\documentclass{eas}
\usepackage{graphicx}
\begin{document}

\TitreGlobal{Mass Profiles and Shapes of Cosmological Structures}

\title{MOND and Cosmology}
\author{R.H. Sanders}\address{Kapteyn Astronomical Institute,
P.O.Box 800, 9700 AV Groningen, The Netherlands}
%
\runningtitle{MOND and Cosmology}
\setcounter{page}{23}
\index{R.H. Sanders}

%
\begin{abstract} 
I review various ideas on  MOND cosmology and structure
formation
beginning with non-relativistic models in analogy with
Newtonian cosmology.  
I discuss relativistic MOND cosmology in the context of
Bekenstein's theory and propose an alternative biscalar effective theory
of MOND in which the acceleration parameter, $a_0$ is identified
with the cosmic time derivative of a matter coupling scalar field
and cosmic CDM appears as scalar field oscillations of the
auxiliary ``coupling strength'' field. 
\end{abstract}

\maketitle
%
\section{General Remarks}

In modified Newtonian dynamics
it is postulated that the true gravitational acceleration, $g$,
is related to the usual Newtonian acceleration, $g_N$, as
$$g\mu(|g|/a_0) = g_N\eqno(1)$$ where $a_0$ is a fixed parameter
with units of acceleration and $\mu(x)$ is a function
interpolating between the MOND regime ($\mu(x)=x$) and the
Newtonian regime ($\mu(x) = 1$) (Milgrom 1983).
This algorithm is arguably more successful in
explaining aspects of galaxy phenomenology than is dark matter
in the context of the CDM
paradigm (Sanders \& McGaugh 2002).  

But MOND, as a theory, is clearly incomplete; it makes no prediction about
cosmology or structure formation.  The fact that $a_o\approx cH_o$ is
suggestive of a cosmological connection, but the
structure of that cosmology is not evident.  It might be expected 
that a hypothesis positing such a radical departure from
Newtonian dynamics (and hence General Relativity) on the scale of
galaxies would result in a highly unconventional cosmology and that this
would be inconsistent with the phenomenological successes of the
standard Big Bang model-- primarily the nucleosynthesis of the
light elements in their observed abundances (Steigman 2003) and the overall
absence of spectral or spatial distortions in the Cosmic Microwave 
Background radiation (Smoot et al. 1992).  Indeed, it
would seem safe to assume that these phenomenological foundations of the
Big Bang are so firm, that this model for the pre-recombination
Universe should be taken as a
requirement on any alternative theory;  i.e., an alternative
theory should not lead to a radically different cosmological scenario for
the early Universe.

Now we all know that MOND was suggested as an alternative to dark matter.
But if MOND is, in some sense, ``true'' this does not mean that 
dark matter is non-existent. 
Indeed, there is compelling astronomical evidence for 
the existence of a cosmic component of pressure-less dark matter (CDM), 
with an abundance in excess of any possible baryonic component. 
This is essentially the same evidence as that supporting
the ``Concordance Model'' ($\Lambda$CDM) for the Universe:

\noindent 1) The overall amplitude of the first two peaks in the angular
power spectrum of the CMB anisotropies is, given an independent
determination of the optical depth to the last scattering surface,
consistent with the presence, at recombination, of dark matter 
potential wells (Page et al. 2003);  
the implied present density of CDM would be about $\Omega_{CDM}\approx 0.25$.

\noindent 2)  The re-brightening of SNIa at $z\ge 1$ (relative to an empty
coasting Universe), implies matter domination over vacuum energy at this 
relatively recent epoch, again at the level of $\Omega_{CDM}\approx 0.25$
(Tonry et al 2003).

Although the evidence may have been overstated (McGaugh 2004),
these two facts imply that any MOND cosmology should 
reproduce or simulate the global effects of cosmic CDM on early 
structure formation and the expansion history of the Universe.
But it would
be inconsistent with MOND if dark matter made a 
dominant contribution to the present 
mass budget of bound gravitational systems--
galaxies and groups of galaxies.
I will return to this point later, but I first review specific
ideas on MOND cosmology.

\section{Primitive (non-relativistic) MOND cosmology:  Modified dynamics
of fluctuations}

Is MOND consistent with FRW world models in the context
of the Cosmological Principle?  Does MOND promote the formation of
the observed range of structure starting from near homogeneity 
at decoupling in a Universe
without CDM?  One might hope to provide answers to these fundamental 
questions by considering the evolution of a uniform sphere in the context
of MOND (Felten 1984, Sanders 1998).  We know that
the Newtonian evolution of a homogeneous sphere expanding against its own 
gravity provides a non-relativistic 
derivation of the Friedmann equations for
the time dependence of the cosmic scale factor. So, does the MONDian evolution
of such an object lead to similar insights into the structure of
a MOND cosmology?  To assert that it does requires two assumptions:

\noindent 
1. The external Universe does not affect the dynamics of a small spherical
piece of the Universe; i.e., there exists an equivalent to the Birkhoff theorem
for the relativistic theory underlying MOND (this is probably not
true).

\noindent 2. The MOND acceleration parameter, $a_0$, is constant with cosmic
time (also a questionable proposition).

The well-known Newtonian equation for the evolution of the radius
$r$ of the sphere is
$$\ddot r = -{{4\pi G r}\over 3} (\rho + 3p). \eqno(2)$$
The fact that acceleration is proportional to radius means
that there exists a critical radius, $r_c=\sqrt{GM/a_0}$, beyond which the
acceleration exceeds $a_0$, so we might expect that on larger
scales the evolution is described by the usual Friedmann equations.
In Friedmann models the critical radius increases as the
dimensionless scale factor: $r_c\propto a(t)^m$ where m= 4 in a 
radiation dominated Universe and m=3 in a matter dominated 
Universe.  Thus, in a MONDian universe we would seem to have
the possibility of Friedmann expansion on the scale of the horizon,
but MOND expansion and re-collapse on smaller scales. 
That is, as soon as the deceleration of a given co-moving region falls
below $a_0$, the dynamical equation becomes
$$\ddot r = -\Bigl[{{4\pi G a_o r}\over 3}(\rho+3p)\Bigr]^{1\over 2}. 
\eqno(3)$$
which leads to the eventual re-collapse of any finite
size region, with larger co-moving regions re-collapsing later.

In such a cosmology the evolution of the early Universe would be
as it is in the standard Big Bang ($r_c$ would be very much
smaller than the relevant Jeans scale). Moreover,
in the present Universe, where this no longer the case,
inhomogeneity on large scale ($\approx 10^{16}$ M$_\odot$) would
seem inevitable.  However, 
there are clear problems in principle with this cosmology.  It is
difficult to reconcile Friedmann expansion in a large volume with
MOND re-collapse about every point within that volume.  If re-collapse
occurs only about selected points, what determines the location of
those seeds for re-collapse.  The problem is that density fluctuations
play no role in this scenario for structure formation;  we might expect
that in a proper MOND cosmology structure would develop from the
field of small density fluctuations as in the standard model of 
gravitational collapse.  

To connect structure formation with density fluctuations one must supplement
the above assumptions with
an additional {\it ansatz}:  the MOND algorithm (eq.\ 1)
should only be applied
to the peculiar accelerations that develop about fluctuations and
not to the overall Hubble flow.  This means that the zeroth order
Hubble flow remains intact; there is no MOND in a homogeneous
Universe.  Such a scenario would seem to be more consistent with the
suggested relativistic theories in which MOND phenomenology results
from a scalar field gradient
that dominates the usual gravity force in the limit of low 
field gradients (Sanders 1997, Bekenstein 2004).
Having said this, the de/acceleration
of the Hubble flow over a particular scale may enter as an external
field-- the so-called ``external field effect'' in which the
internal dynamics of a subsystem is influenced by the presence of a
background acceleration field (Milgrom 1983).

These properties have been realized in a non-relativistic two-field
theory of modified dynamics (Sanders 2001) which is similar to the 
Bekenstein-Milgrom Lagrangian-based theory (Bekenstein \& Milgrom 1984).
The two fields supposedly represent usual gravity
and an anomalous MOND force assumed to act only upon over- or 
under-dense regions. Following the same procedure as in Newtonian
cosmology, I find that the growth equation for small density fluctuations is
non-linear even in the regime where the density fluctuations are small
(this is because MOND is non-linear).  Moreover, as we see in Fig.\ 1
the growth is dramatically rapid where the background acceleration
vanishes; i.e., when the vacuum energy density as described by
a cosmological constant becomes comparable to the matter energy
density.

\begin{figure}[h]
   \centering
\includegraphics[width=9cm]{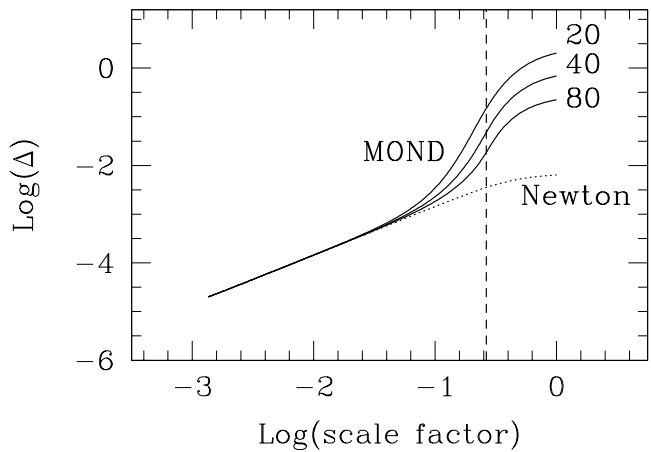}
      \caption{The amplitude of density fluctuations ($\Delta = 
 \delta\rho/\rho$) on various co-moving scales (Mpc) as a function of scale 
factor compared 
to the Newtonian growth in a low-density baryonic Universe 
(Sanders 2001).  The growth rate is particularly 
large where the cosmological term begins to dominate the expansion.}
       \label{figure_mafig}
   \end{figure}
This effect adds a new aspect
to the anthropic argument originally given by Milgrom (1989):  we are
observing the Universe at an epoch where the cosmological term has
recently become dominant because it as at this epoch where structure
can form rapidly.

Nusser (2002) and Knebe \& Gibson (2004) have carried out cosmic
N-body simulations where the MOND formula is applied to the peculiar
accelerations.  Here there is no external field effect of the background
Hubble flow (indeed, it is unclear how such an effect might be realized in
N-body simulations).  Nusser finds that, with the value of $a_0$ determined
from galaxy rotation curves ($\approx 10^{-8}$ cm/s$^2$), structure grows very 
rapidly-- the present
amplitude of density fluctuations would be much larger than observed
($\sigma_8>3$).  If, however, $a_0$ is smaller by about a factor of 10,
then the resulting rate of growth is consistent with standard 
(CDM) theory and the topology of the resulting structure is very
similar to that seen in CDM simulations and actually observed
in large galaxy redshift surveys-- essentially one of filaments and
walls surrounding large voids. 

Basically the rapid growth of structure in these simulations is due
to the absence of the external field effect:  the assumption that
the deceleration of the Hubble flow over a finite size region enters
as a background acceleration field tames this exponential growth.  Without
such an effect, some other mechanism, such as a lower value of $a_0$, must be
invoked-- that is, the assumption that $a_0$ is constant over the
history of the Universe must be relaxed.  This illustrates the
essential limitations of developing a MOND cosmology in the absence
of a relativistic theory. Any one of the assumptions upon which
such a cosmology is based may be wrong, and this would  
obviate the results.  Non-relativistic MOND cosmology, while useful
in getting a broad picture of how a MOND universe may appear and how
various assumptions affect the results, has clearly
reached its limits.

\section{Relativistic MOND cosmology}

A consistent relativistic theory of MOND, such
as TeVeS (Bekenstein 2004), permits derivation of
cosmological models and consideration of structure formation 
without additional assumptions.
In TeVeS, the Friedmann equation is standard, apart from an effectively
variable constant of gravity and additional source terms resulting from
the energy density of the scalar and vector fields.  But in TeVeS as
it now stands, the proposed form of its free function presents a problem
in interpolating between a homogeneous evolving universe and
quasi-static mass concentrations.

The scalar field Lagrangian of TeVeS has the form
$$L_s = {1\over 2} [q^2 \phi_{,\alpha}\phi^{,\alpha} + 2V(q)] \eqno(4)$$
where $\phi$ is the matter-coupling field and $q$ is a non-dynamical 
(or at least not explicitly dynamical) auxiliary field which determines
the strength of that coupling. The free function, $V(q)$, can be viewed
as a potential of this non-dynamical field.
Now, because of the algebraic relation
between $q$ and $\phi_{,\alpha}\phi^{,\alpha}$
(i.e., $q\phi^{,\alpha}\phi^{,\alpha} = V'(q)$), this is really an
aquadratic Lagrangian theory in disguise as discussed by Bekenstein.
That is to say, the scalar field Lagrangian can be written 
$$L_s = {1\over 2} F\Bigl({\phi_{,\alpha}\phi^{,\alpha}}l^2
\Bigr) \eqno(5) $$
where $l$ is a length scale.  Here, the MOND interpolating function
is given by 
$\mu = dF(X)/dX$.  If one is to obtain MOND phenomenology in
the limit of low $|\nabla\phi|$, then $V(q)$ must be chosen such that
$F(X) \rightarrow X^{3/2}$ in the limit of small $X$ as originally 
discussed by Bekenstein \& Milgrom (1984).  However, this obviously cannot be
continued into the cosmological regime where $X<0$ (with the sign convention
adapted here).  This was an early problem for the cosmological extension
of AQUAL and it persists for TeVeS (see Sanders 1986).

Bekenstein chooses to solve this problem by taking $V(q)$ such that
$F(X)$ has two discontinuous branches-- one for cosmology ($X<0$),
and one for quasi-static mass concentrations ($X>0$). Bekenstein 
emphasises
that this choice is tentative and not fundamental to the theory.
This is fortunate because we see an immediate problem here with 
respect to the growth of 
fluctuations in an evolving Universe. How do we deal with the
discontinuity between the cosmological regime and the quasi-static
regime?  It would seem impossible
to follow the evolution of structure, at least to the non-linear
level.  

None-the-less it does seem possible to consider
the linear development of large scale structure in the context of TeVeS
with this somewhat awkward free function.
Skordis et al. (2005) have derived homogeneous
Friedmann-like models and note that the models exhibit the tracking
behaviour characterising some scalar field theories of quintessence:
the relative density in the $\phi$ field attains attractor solutions
in the radiation, matter and $\Lambda$ eras.   They
then consider the evolution of linear
perturbations for a range of the parameters of the theory and find 
reasonable agreement with both the angular power spectrum of the CMB 
fluctuations (out to the second peak) 
and with the observed power spectrum of galaxy density
fluctuations, provided that neutrinos are included at a level
of $\Omega_\nu=0.17$-- near the upper limit permitted by experimental
constraints on the electron neutrino mass.  A critical aspect is
the amplitude of the third peak in the CMB anisotropy 
power-spectrum:  dark matter fluctuations seem to be required to
make this peak as large as observed in earlier CMB experiments (see
also McGaugh 2004).

These calculations demonstrate the 
power of a relativistic
theory like TeVeS, in confronting this range of observed cosmic
phenomena with a theory producing MOND in the quasi-static regime.
However, the present arbitrariness of the free function means that the
results must still be considered as tentative.

\section{A cosmological effective theory of MOND}

One short-coming of TeVeS in its present form is that the MOND
acceleration parameter $a_0$ does not appear to arise in a natural
way but must be inserted by hand.  The near numerical coincidence of
$a_0$ with $cH_0$ remains unexplained.  This coincidence suggests that
the correct theory of MOND may be one in which this characteristic
phenomenology arises only in a cosmological background.  That is to say,
MOND should be described by an effective theory which reflects the
influence of cosmology on local particle dynamics, as originally 
supposed by Milgrom (1983, 2002).  Scalar-tensor theory offers
this possibility  (Dicke 1962).
Moreover, in any theory
of stronger attraction in the limit of weak gradients that also
provides the observed degree of gravitational lensing,  
the scalar field must affect particle motion jointly with the
Einstein metric and an additional vector field (Bekenstein \& Sanders 1994, 
Sanders 1997).  The components of the vector are not
invariant under a Lorentz transformation so the theory will 
inevitably single out the cosmological frame as special.

One possibility for such an effective theory is to explicitly include
dynamics for the auxiliary field $q$ in TeVeS; i.e., write a kinetic term
$q_{,\alpha}q^{,\alpha}$ into the scalar Lagrangian eq.\ 4. 
The theory then becomes a biscalar preferred frame generalisation of 
``phase coupling gravitation'' (PCG), an earlier covariant theory 
also proposed by Bekenstein (1988)
as a basis for MOND. Although PCG, in original form, contains pathologies,
it is attractive in the sense that it permits sensible cosmologies
where the fields at large
distances from mass concentrations naturally approach their cosmological
values (Sanders 1989).
In the biscalar theory, as in PCG, one field $\phi$ couples to matter and the
second field $q$ determines the strength of that coupling.
It is then possible to write down a 
theory in which the MOND phenomenology arises in an evolving Universe
where $a_0$ is identified with $c\dot\phi$ (Sanders 2005).  
Here $a_0$ evolves with
cosmic time in the sense that it was smaller in the past (a factor of
10 smaller at z=10).
\begin{figure}[h]
   \centering
\includegraphics[width=9cm]{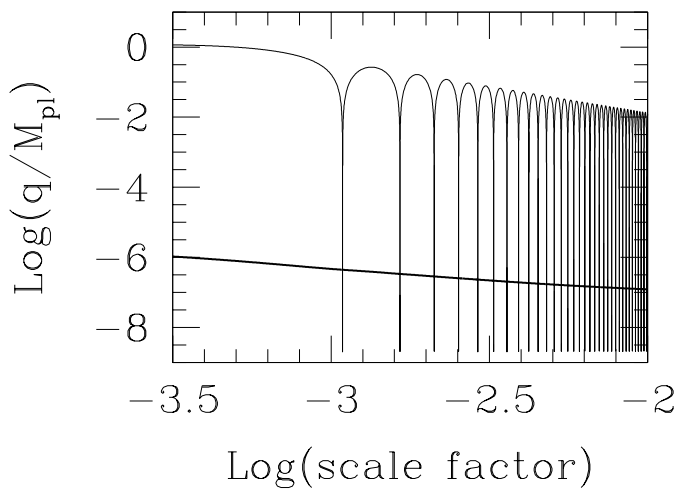}
      \caption{Oscillations that develop in the ``coupling strength
field'' $q$ as it seeks the minimum in the potential well.  These 
can constitute CDM with a large de Broglie wavelength (Sanders 2004)}
       \label{figure_mafig}
   \end{figure}

The cosmology is standard,
but oscillations of the $q$ field inevitably develop as the
field settles to the potential minimum (see Fig.\ 2).  If the bare potential is
quadratic ($V(q) = {1\over 2} A q^2 + B$), these oscillations comprise
cold dark matter, but, depending upon the parameters of the theory,
the de Broglie wavelength of these bosons may be so large that the
the dark matter does not cluster on the scale of galaxies.
So beginning with a theory involving two scalar fields, the matter coupling
field provides MOND phenomenology in the cosmological background, but
oscillations in coupling-strength field provide cosmological dark matter;
an effective theory of MOND produces cosmological CDM for free.

The overall picture is that cosmology is described by a preferred frame
theory with a long range force mediated by a scalar field coupled
to a dynamical vector as well as the gravitational metric.  The fact
that the scalar coupling to matter becomes very weak in the region
of high field gradients protects the solar system from
observable preferred frame effects; i.e.,
the theory becomes effectively identical to General Relativity
in this limit.  The 
outskirts of galaxies would be the transition region between preferred
frame cosmology and a GR dominated local dynamics.  This transition
would be observable as an acceleration-dependent deviation from 
Newtonian dynamics-- MOND.

\end{document}